\documentclass[twocolumn,pre,nobibnotes]{revtex4}
\usepackage{graphicx,amsmath,bm,color,epstopdf,CJK,ulem}

\begin{document}

\title{Clusters and the entropy in opinion dynamics on complex networks}
\author{Wenchen Han}
\affiliation{College of Physics and Electronic Engineering, Sichuan
 Normal University, Chengdu, 610101, China}
\author{Yuee Feng}
\affiliation{College of Arts and Sciences, Jiangsu Aviation Technical College,
 Zhenjiang, 212134, China}
\author{Xiaolan Qian}
\affiliation{School of Media and Engineering, Communication University
of Zhejiang, Hangzhou, 310018, China}
\author{Qihui Yang}
\affiliation{Department of Electrical and Computer Engineering, Kansas State
University, Manhattan, Kansas 66506, USA}
\author{Changwei Huang}\email{cwhuang@gxu.edu.cn}
\affiliation{School of Computer, Electronics and Information, Guangxi University,
Nanning, 530004, China}

\begin{abstract}
In this work, we investigate a heterogeneous population in the modified Hegselmann-Krause
opinion model on complex networks. We introduce the Shannon information entropy
about all relative opinion clusters to characterize the cluster profile in the
final configuration. Independent of network structures, there exists the optimal
stubbornness of one subpopulation for the largest number of clusters and
the highest entropy. Besides, there is the optimal bounded confidence (or
subpopulation ratio) of one subpopulation for the smallest number of clusters
and the lowest entropy.
However, network structures affect cluster profiles indeed. A large
average degree favors consensus for making different networks more similar
with complete graphs. The network size has limited impact on cluster profiles
of heterogeneous populations on scale-free networks but has significant effects upon
those on small-world networks.
\end{abstract}

%\pacs{64.60.ah, 64.60.aq}
\maketitle

\section{Introduction}
The field of multiagent network dynamics, which can reveal the emergence
of collective behavior characteristics based on local individual interaction
rules, has been investigated by researchers from a vast range of disciplines
in recent years \cite{cas09,wang16,gal18}. Agents' opinions can change in time
subject to the interactions between the neighboring agents as well as the global
feedback and external factors.
The opinion dynamics characterized with statistical physics
has been developed in the social opinion research.
Opinion dynamics models are classified
into discrete opinion dynamics models and continuous ones. More specifically,
opinions are modeled as variables, discrete or continuous.
In the discrete case, binary opinion models are most studied in terms of their
analogy with spin systems. Well-known discrete opinion dynamics models include
the voter model \cite{hol75}, the Sznajd model \cite{szn00}, the Galam
majority-rule model \cite{gal02}, and the nonconsensus opinion model
\cite{shao09,yang15}. Among the continuous opinion dynamics models,
the Deffuant-Weisbuch model \cite{def00} and the Hegselmann-Krause (HK)
model \cite{heg02} have attracted much
attention. In the final stationary state, opinion clusters can be
one (consensus), two (polarization), or many (fragmentation).

Opinions are continuous variables in the interval $[0,1]$ in an HK model.
The evolution is treated as a series of discrete time steps. At each time
step, an agent updates her opinion by adopting the average of her
neighbors' opinions and her own. Two agents are neighbors only when
the difference of their opinions is below the threshold, i.e., the bounded confidence.
In a homogeneous population, where agents are with a same set of parameters,
the consensus is the result of agents with a high bounded confidence,
while a fragmentation is caused by a low bounded confidence.
On the other hand, agents with heterogeneous bounded confidences could help agents
to reach consensus \cite{lor09,pin15}. Fu \textit{et al.} found
that agents with high bounded confidence could not contribute to
forging consensus and might lead to fragmentation \cite{fu15}. While
Han \textit{et al.} \cite{han19} found whether agents would reach consensus
depended upon not only the bounded confidence but also the stubbornness,
which measured the extent of an agent's insistence on her own opinion
\cite{gha14,ols09,cha17,chen19}.

The network structure provides a background for the opinion dynamics
system, in which nodes stand for agents and links connecting the nodes
represent the possible interactions between agents. The network
structure can be a complete graph, a random network \cite{ero60},
a scale-free network \cite{bar99}, or a small-world
network \cite{watt98}. However, it is known that network
structures affect dynamical processes on networks \cite{port16,meng18,ask15}.
Results of complete graphs and square lattices are similar for
large bounded confidence values, except for remaining a few
extreme opinions \cite{wei02}. Jalili studied the social power for
the consensus in opinion dynamics on complex networks, found that
giving the social power to nodes in small-world networks cannot
significantly affect the consensus while hubs with social power in
scale-free networks can \cite{jal13}. Even for small-world networks,
directed networks and bi-directed networks show different effects
on opinion dynamics \cite{gan10}. Bagnoli and Rechtman studied a
society with conformist and reasonable contrarian agents, and they
found systems on small-world networks and scale-free networks showed
some similarities due to the long-range connectivity \cite{bag13,bag15}.

In this work, we considered a heterogeneous population on different
complex networks, where agents in a same subpopulation shared a same
set of parameters (i.e., bounded confidence and stubbornness) and different
subpopulations were with different sets of parameters. We paid our attention
to the cluster profile while varying one of the parameters in this heterogeneous
population. The cluster profile contained the information about the number of
opinion clusters and the relative sizes of all clusters, and it implied
the Shannon information entropy about the sizes of clusters.
We found that the cluster profile was the trade-off of competition effects
of agent's parameters. Furthermore, we also investigated the effect of the
average degree of networks and the effect of the network size on
the cluster profile. A large average degree made networks behave like
a complete graph, supporting fewer clusters, a larger size of the largest cluster,
and a low entropy. Besides, the network size had less impact on
cluster profiles of populations on scale-free networks but significant impacts
on those on small-world networks, due to the strongly connected local structures.

\section{Model}

We consider a heterogeneous population composed of $N$ agents on complex networks.
Agent $i$ is placed on the node $i$ in the network. $A_{ij}$ is an
element of the adjacency matrix for the network structure.
$A_{ij}=A_{ji}=1$ denotes node $i$ and node $j$ are connected
and that agent $i$ and agent $j$ may be neighbors.
$A_{ij}=A_{ji}=0$ indicates that node $i$ and node $j$ are
disconnected and that agent $i$ and agent $j$ cannot
be neighbors. Additionally, self-edges are avoided ($A_{ii}=0$).
The opinion of agent $i$ at discrete time $t$ is described by
a continuous variable $x_i(t)$, satisfying $0\le x_i(t)\le1$.
The evolution of the opinion of agent $i$ follows the updating rule
\begin{equation}
x_i(t+1)=\begin{cases} \alpha_ix_i(t)+\frac{(1-\alpha_i)}{||N_i(t)||}
\sum\limits_{j\in N_i(t)}{x_j(t)},&||N_i(t)||>0,\\
x_i(t),&||N_i(t)||=0. \end{cases}
\end{equation}
$\alpha_i$, which is in the interval $[0,1]$, describes the stubbornness
of agent $i$, suggesting the extent of agent $i$ insisting on her own opinion.
$\alpha_i=0$ means agent $i$ is a
conformist, just following other neighbors' opinions, while agent $i$
with $\alpha_i=1$ is a zealot, only sticking to her own opinion and
neglecting others \cite{wang16b}. Note that agent $i$ is more stubborn than agent $j$
when $\alpha_i$ is larger than $\alpha_j$. $N_i(t)$ for agent $i$ at time $t$
is the set of neighbor agents,
whose opinions satisfy $|x_j(t)-x_i(t)|<\sigma_i$ and $A_{ij}=1$.
$\parallel N_i(t)\parallel$ is the cardinality of $N_i(t)$.
Bounded confidence $\sigma$ reflects the psychological concept
of selective exposure, which refers to an individual's tendency to
favor information that supports her opinion while neglecting conflicting
arguments, with $\sigma\in(0,1]$. We name agents with high
$\sigma$ as open-minded ones and agents with low $\sigma$ as
close-minded ones.

In this work, the total population is composed of $M$ subpopulations.
Each subpopulation with $\rho_lN$ agents, with the ratio $\rho_l\in[0,1]$
and $\sum^{l=M}_{l=1}{\rho_l=1}$, takes the
bounded confidence $\sigma_l$ and the stubbornness $\alpha_l$. The
subpopulation with $\sigma_l$ and $\alpha_l$ is denoted as $C_l$ for
convenience. When opinions of all agents stop evolving,
we say the system reaches its steady state. Then we monitor the number
of opinion clusters, $N_c$, where connecting agents within a same cluster
share a same opinion. Unconnected agents with a same opinion
belong to different clusters. So it is with connected agents with different opinions.
$S_i$ is the relative size of an opinion cluster, namely
the ratio between the number of agents in the cluster and the number
of agents in the system. The relative size of all opinion clusters
can be sorted in a descending order, i.e., $1\ge S_1\ge S_2\ge \cdots
\ge S_{N_c}>0$. Additionally, $\{S_1,S_2,\cdots,S_{N_c}\}$ is an opinion
cluster profile.
We introduce the Shannon information entropy $H$ \cite{shan48} as
\begin{equation}
H=-\sum^{N_c}_{i=1}{S_i\log_2 S_i}.
\end{equation}
The Shannon information entropy, analogy with Boltzmann Entropy \cite{bag13,bag15},
describes the information gain of a certain opinion cluster profile.
When $N_c=1$ for the existence of only one cluster and $S_1=1$, no information
can be gained ($H=0$). If the opinions are scattered (e.g., $N_c=N$ and $S_i=1/N$
for $i=1,2,\cdots,N$), much information can be gained ($H=\log_2(N)$). For similar
cluster sizes, a cluster profile with more clusters corresponds to a higher $H$.
If two opinion cluster profiles share a same number of clusters, the profile
composed of less heterogeneous clusters, where some clusters are with similar sizes,
are with a higher $H$. In this work, we consider the connecting structure
between agents as complex networks, including random networks (RNs),
scale-free networks (SFNs), and small-world networks (SWNs)
with rewiring possibility $0.1$.
All the networks are connected and have $N$ nodes and the $NK/2$ links,
where $K$ is the average degree. We consider the population size $N=1000$
and the average degree $K=20$ unless specified. The simulation results are
averaged over $100$ realizations with uniform random initial conditions in
opinions. The opinions of agents are updated synchronously.

\section{Simulation results}

In a homogeneous population ($M=1$), the stubbornness $\alpha$ only affects
the transition time when the bounded confidence $\sigma$ is fixed \cite{jan07}.
A larger $\alpha$ is for a longer transition process. Increasing $\sigma$ of agents always
leads to a smaller number of opinion clusters $N_c$ and a larger size of the largest
opinion cluster $S_1$ when the bounded confidence is smaller than the threshold
($\sigma_c=0.25$). Beyond $\sigma_c$, only one opinion cluster exists.
It is also known that the dependence of $N_c$ (or $S_1$) on
$\sigma$ is caused by the competition effects of agents' bounded confidences
and stubbornness in a heterogeneous population ($M\ge2$) on complete
graphs \cite{han19}.

In the following, we will report the results in a heterogeneous
opinion population on complex networks (RNs, SFNs, SWNs) and pay attention to
the cluster profile in the final configuration, which can be depicted
by $N_c$, ${S_i}, i=1,2,\cdots,N_c$, and $H$. We consider a
heterogeneous population just composed of two subpopulations $M=2$,
where $\rho_1N$ agents are with $\sigma_1$ and $\alpha_1$, $\rho_2N$ agents
are with $\sigma_2$ and $\alpha_2$, and $\rho_1+\rho_2=1$.

\begin{figure}
\begin{center}
\includegraphics[width=3.4in]{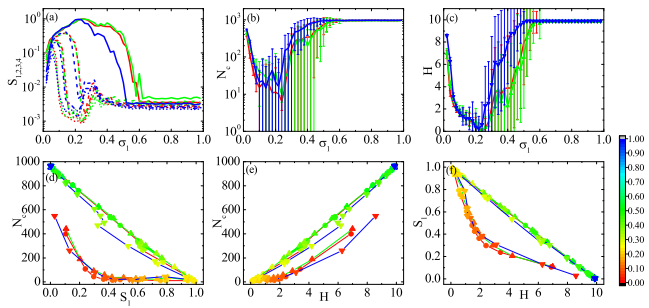}
\end{center}
\caption{\label{fig_1} The dependent relations for varying $\sigma_1$
when $\rho_1=0.9$, $\alpha_1=0.8$, $\rho_2=0.1$, $\alpha_2=0$,
and $\sigma_2=0.04$. The red line/circle dots for random networks,
the green line/up-triangles dots for scale-free networks,
the blue line/down-triangles dots for small-world networks.
(a) the dependence of four largest clusters
$S_{1,2,3,4}$ (solid, dash, dash-dot, short dash lines) on $\sigma_1$.
(b) the dependence of the number of clusters $N_c$ on $\sigma_1$.
(c) the dependence of the Shannon information entropy $H$ on $\sigma_1$.
(d) the relation of $N_c$ against $S_1$. (e) the relation of $N_c$ against $H$.
(f) the relation of $S_1$ against $H$. In panels (d-f), the colors of dots
for the values of $\sigma_1$ linearly.}
\end{figure}

In Fig.~\ref{fig_1}, $C_1$ agents are more stubborn than close-minded
$C_2$ agents ($\rho_1=0.9$, $\alpha_1=0.8$, $\rho_2=0.1$, $\alpha_2=0.0$,
and $\sigma_2=0.04$), where $C_2$ subpopulation tends to form about $10$
clusters when only $C_2$ agents are on a complete graph. In Fig.~\ref{fig_1} (a),
increasing the bounded confidence of $C_1$ agents makes $S_1$ first increase,
then decrease, and remain constant finally ($\sigma_1>0.6$), while it
makes $N_c$ decrease, increase, then remain constant, as shown in
Fig.~\ref{fig_1} (b). Interestingly, when the $C_1$ agents are much
open-minded (e.g., $\sigma_1=0.8$), all clusters are small and the size
of the largest opinion cluster is even smaller than that when $C_1$
agents are close-minded (e.g., $\sigma_1=0.1$). A high $\sigma$ leads
to fragmentation \cite{fu15,han19}. However, in this case, this is caused
by not only the competition effect between $\sigma$ and $\alpha$
but also the connection constrains of network structures, which can be
verified in the following discussion.
However, the dependence of $S_1$ on $\sigma_1$ and the dependence
of $N_c$ on $\sigma_1$ are not simply negative correlated, as shown
in Fig.~\ref{fig_1} (d). For example, $S_1$ increases while $N_c$
almost remains when $0.1\le\sigma_1\le0.2$. This means that only
considering $S_1$ and neglecting the evolution of other opinion clusters
cannot explain the behavior of $N_c$, which counts up all clusters.
That's because $S_1$ keeps increasing along with the decreases of $S_4$, $S_3$,
and $S_2$ when $N_c$ almost remains. The Shannon information
entropy is a good measure for taking all clusters into account. It encodes
not only the number but also the heterogeneity of all opinion clusters.
In Fig.~\ref{fig_1} (c), $H$ decreases when $S_1$
increases along with $S_4$, $S_3$, $S_2$ decreasing for $\sigma_1\in(0.1,0.2)$.
$H$ shows some similar features but not exactly the same with $N_c$,
which is not a strict linear relationship as shown in Fig.~\ref{fig_1} (e).
Figure~\ref{fig_1} (f) shows the relation of $S_1$ against $H$ is a
roughly negative linear one, suggesting that a same $H$ corresponds to
different $S_1$ and vice versa.
Figure~\ref{fig_1} (d-f) suggests
that one of $H$, $N_c$, and $S_1$ corresponds to different cluster profiles
and that only one of $H$, $N_c$, and $S_1$ is not sufficient to describe a cluster profile.
Additionally, Fig.~\ref{fig_1} shows cluster profiles on
different networks share many characteristics.

\begin{figure}
\begin{center}
\includegraphics[width=3.4in]{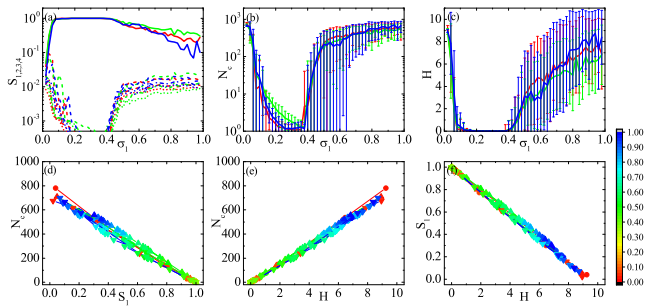}
\end{center}
\caption{\label{fig_2} The dependent relations for varying $\sigma_1$
when $\rho_1=0.9$, $\alpha_1=0$, $\rho_2=0.1$, $\alpha_2=0.8$,
and $\sigma_2=0.3$. The red line/circle dots for random networks,
the green line/up-triangles dots for scale-free networks,
the blue line/down-triangles dots for small-world networks.
(a) the dependence of four largest clusters
$S_{1,2,3,4}$ (solid, dash, dash-dot, short dash lines) on $\sigma_1$.
(b) the dependence of the number of
clusters $N_c$ on $\sigma_1$. (c) the dependence of the Shannon
information entropy $H$ on $\sigma_1$. (d) the relation of $N_c$
against $S_1$. (e) the relation of $N_c$ against $H$. (f) the relation
of $S_1$ against $H$. In panels (d-f), the colors of dots for the values
of $\sigma_1$ linearly.}
\end{figure}

Figure~\ref{fig_2} shows the results of $C_1$ agents ($\alpha_1=0.0$) with
stubborn but open-minded $C_2$ agents ($\alpha_2=0.8$ and $\sigma_2=0.3$),
where $C_2$ agents can achieve consensus when only $C_2$ agents exists on
a complete graph.
The dependence of $S_1$, $N_c$, and $H$ on $\sigma_1$ is qualitatively the same
as that of the population in Fig.~\ref{fig_1} regardless of different values
of $\alpha_1$, $\alpha_2$, and $\sigma_2$. However, it is a little different
from the result when a heterogeneous population sitting on complete
graphs \cite{han19}, where increasing $\sigma_1$ always leads to consensus
when the other subpopulation agents are open-minded. A steep decrease in $H$
suggests a quick process of clusters merging when $\sigma_1\in(0.02,0.1)$,
presented by the largest cluster increasing with the vanishing of small
clusters. However, further increasing $\sigma_1$ even leads to a smaller $S_1$,
a larger $N_c$, and a higher $H$ in surprise, when $\sigma_1>0.4$,
evolving on complex networks. These results are constrained by
connections of network structures.
Different from what is shown in Fig.~\ref{fig_1},
$N_c$ against $S_1$  (Fig.~\ref{fig_2} (d)) and $S_1$ against $H$
(Fig.~\ref{fig_2} (f)) exhibit highly negative linear relations and
$N_c$ against $H$ (Fig.~\ref{fig_2} (e)) is with a highly linear relation.
These high correlations imply that one of $H$, $N_c$, and $S_1$ is enough to
describe a cluster profile, in this case. Moreover, the dependent relations
show no difference between populations on networks.

\begin{figure}
\begin{center}
\includegraphics[width=3.4in]{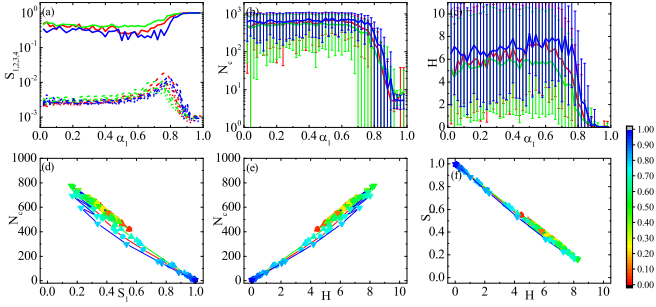}
\end{center}
\caption{\label{fig_3} The dependent relations for varying $\alpha_1$
when $\rho_1=0.9$, $\sigma_1=0.3$, $\rho_2=0.1$, $\alpha_2=0$,
and $\sigma_2=0.04$. The red line/circle dots for random networks,
the green line/up-triangles dots for scale-free networks,
the blue line/down-triangles dots for small-world networks.
(a) the dependence of four largest clusters
$S_{1,2,3,4}$ (solid, dash, dash-dot, short dash lines) on $\sigma_1$.
(b) the dependence of the number of
clusters $N_c$ on $\sigma_1$. (c) the dependence of the Shannon
information entropy $H$ on $\sigma_1$. (d) the relation of $N_c$
against $S_1$. (e) the relation of $N_c$ against $H$. (f) the relation
of $S_1$ against $H$. In panels (d-f), the colors of dots for the values
of $\alpha_1$ linearly.}
\end{figure}

By varying $\alpha_1$, cluster profiles for open-minded agents ($\sigma_1=0.3$)
interacting with not stubborn and close-minded agents ($\sigma_2=0.04$,
$\alpha_2=0$) are shown in Fig.~\ref{fig_3}. Slightly different from what
happens on complete graphs, where increasing the stubbornness of open-minded
agents always leads to smaller clusters and a larger largest cluster \cite{han19},
on complex networks $S_1$ decreases then increases while $N_c$ and $H$
increases then decreases when $\alpha_1$ increases.
When $C_1$ agents are not so stubborn, the decrease of $S_1$ is caused by
the local equilibrium, where $C_1$ agents quickly adjust their opinions to
the average opinion of their neighbors. While $C_1$ are very stubborn,
they wait their neighbors to adjust their opinions, which favors the
cluster merging process and leads to a larger largest clusters.
We have also checked the results of some other heterogeneous population.
The (negative) linear relation of $N_c$ against $H$
($N_c$ against $S_1$ or $S_1$ against $H$) is robust.
It can be inferred that one of $H$, $N_c$, and $S_1$ is sufficient for describing
cluster profiles when varying the stubbornness of one subpopulation.
Furthermore, the dependent relations on $\alpha_1$ is almost independent
of network structures.

\begin{figure}
\begin{center}
\includegraphics[width=3.4in]{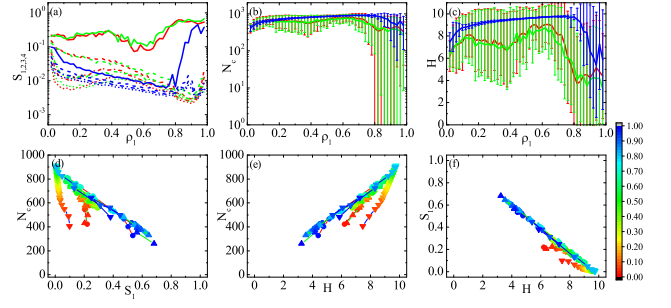}
\end{center}
\caption{\label{fig_4} The dependent relations for varying $\rho_1$
when $\alpha_1=0$, $\sigma_1=0.3$, $\rho_2=1-\rho_1$, $\alpha_2=0.8$,
and $\sigma_2=0.04$. The red line/circle dots for random networks,
the green line/up-triangles dots for scale-free networks,
the blue line/down-triangles dots for small-world networks.
(a) the dependence of four largest clusters
$S_{1,2,3,4}$ (solid, dash, dash-dot, short dash lines) on $\sigma_1$.
(b) the dependence of the number of
clusters $N_c$ on $\sigma_1$. (c) the dependence of the Shannon
information entropy $H$ on $\sigma_1$. (d) the relation of $N_c$
against $S_1$. (e) the relation of $N_c$ against $H$. (f) the relation
of $S_1$ against $H$. In panels (d-f), the colors of dots for the values
of $\rho_1$ linearly.}
\end{figure}

The ratios of subpopulations affect the influence of the dependent relations,
as shown in Fig.~\ref{fig_4}, where the open-minded subpopulation $C_1$
($\alpha_1=0$, $\sigma_1=0.3$) is with a more stubborn close-minded subpopulation
$C_2$ ($\alpha_1=0.8$, $\sigma_1=0.04$). In this case, cluster profiles
on different complex networks share some characteristics,
while cluster profiles of populations on SWNs show some differences
from those on RNs or SFNs. In Fig.~\ref{fig_4} (a), when $\rho_1$ is rather small,
the four largest clusters are with similar sizes in spite of network structures.
Increasing the ratio of $C_1$ subpopulation $\rho_1$ when $\rho_1<0.8$,
the size of the largest cluster $S_1$ almost remains along with other clusters
shrinking on RNs or SFNs, while the largest cluster together with
other clusters shrinks on SWNs. When $\rho_1$ further increases,
$S_1$ continues to increase and it increases to around $0.5$.
Furthermore, the number of clusters $N_c$ when populations evolve on SWNs
is comparable larger than that on RNs or SFNs in Fig.~\ref{fig_4} (b).
So it is with the entropy $H$ in Fig.~\ref{fig_4} (c).
Although the roughly linear dependent relations remain, differences between
populations on SWNs and those on RNs or SFNs are more obvious, as shown
in Fig.\ref{fig_4} (d-f). On SWNs, the population will show more clusters
with smaller $S_1$, higher $H$, and larger $N_c$.
This is caused by the strongly connected local structures,
causing local equilibrium, in SWNs.

The results above show the effects of parameters of agents on cluster
profiles of heterogeneous populations. It is convinced that cluster profiles
are the trade-off of competition between all parameters \cite{han19}
and that features of cluster profiles are almost independent of network
structures when only one parameter of agents varies.
However, network structures should have some effects on cluster profiles,
as known in Ref.~\cite{port16,meng18,ask15} and shown in Fig.~\ref{fig_4}.
In the following, two parameters of network structures, the average
degree $K$ and the network size $N$, are concerned.

\begin{figure}
\begin{center}
\includegraphics[width=3.4in]{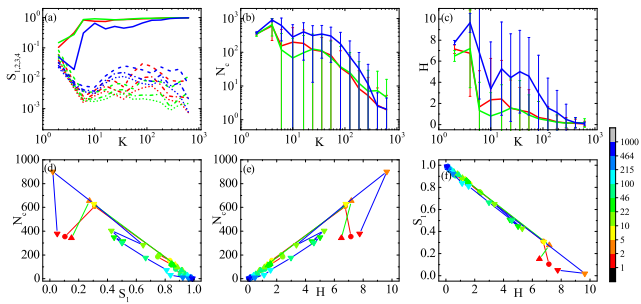}
\end{center}
\caption{\label{fig_5} The dependent relations for varying $K$
when $\rho_1=0.9$, $\alpha_1=0.8$, $\sigma_1=0.3$, $\rho_2=0.1$,
$\alpha_2=0$, and $\sigma_2=0.04$, with the population size $N=1000$.
The red line/circle dots for random networks,
the green line/up-triangles dots for scale-free networks,
the blue line/down-triangles dots for small-world networks.
(a) the dependence of four largest clusters $S_{1,2,3,4}$ (solid, dash,
dash-dot, short dash lines) on $\sigma_1$. (b) the dependence of the number
of clusters $N_c$ on $\sigma_1$. (c) the dependence of the Shannon
information entropy $H$ on $\sigma_1$. (d) the relation of $N_c$
against $S_1$. (e) the relation of $N_c$ against $H$. (f) the relation
of $S_1$ against $H$. In panels (d-f), the colors of dots for the values
of $K$ logarithmically.}
\end{figure}

We first study the influence of the average degree $K$ on these
dependent relations. $S_1$ increases while $N_c$ and $H$ decrease
when $K$ increases, as shown in Fig.~\ref{fig_5} (a-c).
The difference between SWNs and RNs or SFNs still remains when
$K$ is small, but it is vanishing when $K$ is rather large.
It should be noticed that when $K$ increases with $N$ fixed the
difference between different network structures vanishes.
When the network connectivity is high ($K/N\rightarrow 1$), all
network structures will show the characteristics of complete graphs.
The results in Fig.~\ref{fig_5} (d-f) confirm this, with all dot-lines for different
networks converging to a same point, which almost corresponds to the
result for the population evolving on complete graphs.

\begin{figure}
\begin{center}
\includegraphics[width=3.4in]{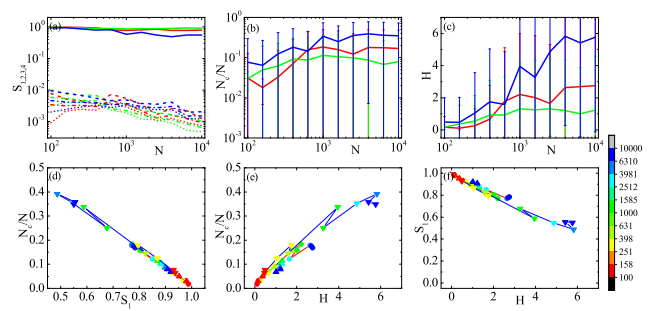}
\end{center}
\caption{\label{fig_6} The dependent relations for varying $N$
when $\rho_1=0.9$, $\alpha_1=0.8$, $\sigma_1=0.3$, $\rho_2=0.1$,
$\alpha_2=0$, and $\sigma_2=0.04$, with the average degree $K=20$.
The red line/circle dots for random networks,
the green line/up-triangles dots for scale-free networks,
the blue line/down-triangles dots for small-world networks.
(a) the dependence of four largest clusters $S_{1,2,3,4}$ (solid, dash,
dash-dot, short dash lines) on $\sigma_1$. (b) the dependence of the relative number
of clusters $N_c/N$ on $\sigma_1$. (c) the dependence of the Shannon
information entropy $H$ on $\sigma_1$. (d) the relation of $N_c/N$
against $S_1$. (e) the relation of $N_c/N$ against $H$. (f) the relation
of $S_1$ against $H$. In panels (d-f), the colors of dots for the values
of $N$ logarithmically.}
\end{figure}

We also investigate the effects of the network size $N$.
In this case, we monitor the relative number of clusters $N_c/N$
in place of the absolute number of clusters $N_c$. Despite of network
structures, $S_1$ is decreasing while $N_c/N$ and $H$ are increasing
when $N$ increases. Thanks to the strongly connected local structures,
the variation of $H$ on SWNs is more obvious than that on RNs or SFNs.
Although the linear relations are robust, the differences between the networks
are shown in Fig.~\ref{fig_6} (d-f) directly.
The increase of $N$ shows the least effect on SFNs and the most effects on
SWNs, which suggests that effects of strongly connected local structure in SWNs
are strongly affected by $N$.

\section{Conclusion and discussion}

In this work, we studied cluster profiles of a heterogeneous population
obeying the modified Hegselmann-Krause opinion dynamic rule on complex networks.
The cluster profile could be described by not only the number of clusters,
but also the relative size of the largest cluster, as well as the Shannon
information entropy about relative sizes of all clusters.
We found the cluster profile was the trade-off of competition between
all parameters of this heterogeneous population, including the stubbornness,
the bounded confidence, and the subpopulation ratio.
Except that even if agents in another subpopulation were open-minded,
increasing the bounded confidence could lead to more clusters and a
smaller size of the largest cluster on complex networks, which was
quite similar with that on complete graphs \cite{han19}.
Optimal bounded confidence of one subpopulation existed for the
fewest clusters, the largest size of largest cluster,
and the lowest information gain. While increasing the stubbornness of
one subpopulation led to the most clusters and the highest entropy.
Increasing the ratio of open-minded agents was not always beneficial
to the fewest clusters and the lowest entropy.
Moreover, we found that the dependent relations between $H$, $N_c$, and $S_1$
were almost independent of network structures when only one of parameters varied,
including bounded confidence, stubbornness, and subpopulation ratio.

Furthermore, we also investigated effects of parameters of
network structures on cluster profiles, such as the average degree and
the network size. The difference between network structures remained
when the average degree was small, while it vanished when the
average degree was large enough. The large average degree made networks
similar with a complete graph, where cluster profiles on it had a
larger $S_1$, a smaller $N_c$, and a lower $H$. As for the network size,
cluster profiles of heterogeneous populations on scale-free networks
shared many similar features, while those on small-world networks
showed more differences, and those on random networks were in the medium.
These different results between different network structures were
caused by the strongly connected local structures in small-world networks,
hubs-leaf structures in scale-free networks, and random connection
structure in random networks. These two findings showed that network structures
could affect the evolution of opinion dynamics indeed.

Our work sheds light upon the cluster profile of a heterogeneous population
on complex networks. Cluster profiles on different networks have many
similar features but also display some differences. However, the conclusion in
this work is just from one perspective. It is worthy studying the mechanisms
to forge consensus from other perspectives and in detail.
For example, it may be interesting to study a heterogeneous population
on some special networks, such as networks with symmetry \cite{kli19}.

\end{document}